\documentclass[letters,fleqn,usenatbib]{mnras}

\usepackage{newtxtext,newtxmath}

\usepackage[T1]{fontenc}
\usepackage{ae,aecompl}

\usepackage{graphicx}	
\usepackage{amsmath}	
\usepackage{amssymb}	

\title[Inner bars also buckle]{Inner bars also buckle. The MUSE TIMER view of the double-barred galaxy NGC~1291.
}

\author[J. M\'endez-Abreu]{
J. M\'endez-Abreu,$^{1,2}$\thanks{E-mail: jairomendezabreu@gmail.com}
A. de Lorenzo-C\'aceres,$^{1,2}$
D. A. Gadotti,$^{3}$
F. Fragkoudi,$^{4}$
\newauthor G. van de Ven,$^{3}$
J. Falc\'on-Barroso,$^{1,2}$
R. Leaman,$^{5}$
I. P\'erez,$^{6,7}$
M. Querejeta,$^{3,8}$
\newauthor P. S\'anchez-Blazquez,$^{9}$
M. Seidel$^{10,11}$
\\
$^{1}$Instituto de Astrof\'isica de Canarias, Calle V\'ia L\'actea s/n, E-38205 La Laguna, Tenerife, Spain\\
$^{2}$Departamento de Astrof\'isica, Universidad de La Laguna, E-38200 La Laguna, Tenerife, Spain\\
$^{3}$European Southern Observatory, Karl-Schwarzschild-Str. 2, D-85748 Garching bei Munchen, Germany\\
$^{4}$Max-Planck-Institut f\"ur Astrophysik, Karl-Schwarzschild-Str. 1, D-85748 Garching bei M\"unchen, Germany\\
$^{5}$Max-Planck Institut f\"ur Astronomie, K\"onigstuhl 17, D-69117 Heidelberg, Germany\\
$^{6}$Departamento de F\'isica Te\'orica y del Cosmos, Universidad de Granada, Facultad de Ciencias (Edificio Mecenas), E-18071, Granada, Spain\\
$^{7}$Instituto Universitario Carlos I de F\'isica Te\'orica y Computacional, Universidad de Granada, E-18071 Granada, Spain\\
$^{8}$Observatorio Astron{\'o}mico Nacional (IGN), Calle Alfonso XII 3, E-28014 Madrid, Spain\\
$^{9}$Departamento de F\'isica Te\'orica, Universidad Aut\'onoma de Madrid, E-28049 Cantoblanco, Spain\\
$^{10}$Caltech-IPAC, Spitzer Science Center, 1200 E. California Blvd., Pasadena, CA 91125, USA\\
$^{11}$The Observatories of the Carnegie Institution for Science, 813 Santa Barbara St., Pasadena, CA 91101, USA\\
}

\date{Accepted XXX. Received YYY; in original form ZZZ}

\pubyear{2018}

\begin{document}
\label{firstpage}
\pagerange{\pageref{firstpage}--\pageref{lastpage}}
\maketitle

\begin{abstract}
Double bars are thought to be important features for secular evolution
in the  central regions of galaxies.   However, observational evidence
about their  origin and evolution is  still scarce.  We report  on the
discovery  of the  first Box-Peanut  (B/P) structure  in an  inner bar
detected in  the face-on galaxy  NGC~1291.  We use the  integral field
data obtained from the MUSE spectrograph within the TIMER project. The
B/P structure is detected as bi-symmetric minima of the $h_4$
moment of the line-of-sight velocity distribution along the major axis
of  the  inner  bar,  as expected  from  numerical  simulations.   Our
observations  demonstrate  that  inner   bars  can  follow  a  similar
evolutionary path  as outer  bars, undergoing  buckling instabilities.
They  also suggest  that inner  bars are  long-lived structures,  thus
imposing tight constraints to their possible formation mechanisms.
\end{abstract}




\section{Introduction}
\label{sec:intro}

\citet{devaucouleurs75}   described  a   rarity  in   the  center   of
NGC~1291. He detected, for the first  time, an inner bar which follows
the same lens-bar-nucleus pattern of the outer bar.  Nowadays, we know
that  bars within  bars are  not an  oddity, with  recent observations
suggesting that  $\sim30$\% of all  barred galaxies host an  inner bar
\citep{erwin04,erwin11,buta15}.  The  importance of inner bars  is not
restricted to their high incidence.  In particular, they are though to
be an efficient  mechanism for transporting gas to  the galaxy central
regions,     possibly      fueling     active      galactic     nuclei
\citep[AGN;][]{shlosman90,maciejewski02}  and affecting  the formation
of                new                stellar                structures
\citep{delorenzocaceres12,delorenzocaceres13}.  Still, little is known
about  the origin  of inner  bars and  now, 40  years later,  NGC~1291
strikes  again  providing  a  new  piece of  evidence  to  unveil  the
formation of the central regions of galaxies.

The current scenarios leading to the  build up of an inner stellar bar
include: i) gas inflow through the outer bar \citep{friedlimartinet93}
induces the  formation of an inner  gaseous bar that then  forms stars
\citep{heller01,shlosmanheller02,englmaiershlosman04},   or  ii)   gas
inflow  through the  outer bar  creates an  inner stellar  disc, which
becomes   dynamically   cold,  and   forms   an   inner  stellar   bar
\citep{wozniak15}.  The  latter scenario has  also been worked  out in
dissipationless  simulations where  a dynamically  cold stellar  inner
disc    is   imposed    at   the    beginning   of    the   simulation
\citep{rautiainensalo99,rautiainen02,debattistashen07,shendebattista09,sahamaciejewski13,du15}.
Besides the  different origins,  the most striking  difference between
these models  is the fate  of the inner  bar.  In the  current models,
stellar inner  bars built up  of previously formed gaseous  inner bars
are short-lived, lasting only a few galaxy rotations.  Similarly, past
models of inner bar formation where  the inner stellar disc was formed
from gas inflow  along the outer bar, proved to  be short-lived due to
the  destructive effect  of the  central mass  concentration. However,
\citet{wozniak15}   recently   challenged   this  picture   with   new
simulations showing  inner bars could  be long-lived. Therefore  it is
crucial  to  find  not  only  new  inner  bars,  but  also  a  way  to
observationally quantify  their lifetime,  in order to  help constrain
these theoretical scenarios further.

Unlike inner  bars, the so-called galaxy  outer (or main) bars  can be
easily formed in numerical simulations once a dynamically cold disc is
settled  \citep[see][for  a  review]{sellwood14}.  The  formation  and
evolution  of these  bars can  be  followed with  exquisite detail  in
numerical  simulations,  and  they   generally  behave  as  long-lived
structures
\citep{combes90,debattistasellwood00,athanassoula03,kraljic12},    but
see  \citet{bournaud05}. A  crucial epoch  during  the life  of a  bar
occurs when its central region  grows in the vertical direction during
the buckling phase.   During this phase, generally  ocurring after the
bar is completely formed and lasting  about $1-2$ Gyr, the bar creates
a    box/peanut   (B/P)    structure    in    its   central    regions
\citep{combessanders81,athanassoula83,martinezvalpuesta06}.   The  B/P
structures are  readily recognisable in edge-on  galaxies, where their
characteristic    shapes    stand    out    of    the    disc    plane
\citep{lutticke00,yoshino15,ciamburgraham16}.   The high  incidence of
B/P  structures provides  observational evidence  that bars  generally
survive for, at least, several  galaxy rotations thus suggesting they
are long-lived structures  with a minimum age since  their assembly of
$\sim$2-3 Gyrs.

The detection of  B/P structures in low-inclination galaxies  is a far
more  difficult  task  due  to  projection effects,  but  it  is  also
extremely rewarding  since it allows  for a direct  comparison between
the  bar  and  the  B/P   properties.   Three  main  diagnostics  (two
photometric  and one  kinematic) have  been proposed  to identify  B/P
structures in non edge-on galaxies: i) barlenses; this method is based
on detecting  a central, close  to circular, isophotal contour  in the
central region of  the galaxy that it is associated  with the presence
of            a           projected            B/P           structure
\citep{laurikainen11,athanassoula15,laurikainensalo17};    ii)    boxy
center+spurs; this is similar to  the previous criterion, but concerns
more  inclined galaxies  and  it is  based on  the  detection of  boxy
isophotes in the galaxy central region followed by narrower and offset
(with respect  to the  bar major  axis) isophotes  in the  outer parts
\citep[spurs;][]{erwindebattista13,erwindebattista17};  iii) kinematic
features in face-on galaxies; the detection of symmetric double minima
of the fourth-order ($h_4$)  Gauss-Hermite moment of the line-of-sight
velocity distribution (LOSVD)  along the major axis of  the bar, which
was  shown to  be  a signature  of  the presence  of  a B/P  structure
\citep{debattista05,mendezabreu08b,mendezabreu14}

In this letter  we show the first  detection of a B/P  structure in an
inner bar,  namely that of  NGC~1291.  We use the  high-quality stellar
kinematics provided by  the MUSE spectrograph within  the TIMER project
(D. A. Gadotti et al. 2018, submitted, hereafter Paper I) to trace the
presence of symmetric double $h_4$ minima  along the major axis of the
inner bar  of NGC~1291.  In Sect.~\ref{sec:structures}  we describe the
main  stellar   structures  present  in  this   double-barred  galaxy.
Sect.~\ref{sec:observations}  describes  the observation  and  stellar
kinematic     measurements.     The     results    are     shown    in
Sect.~\ref{sec:results} along with  the corresponding discussion.  Our
conclusions are summarised in Sect.~\ref{sec:conclusions}.

\section{NGC~1291: properties and stellar structures}
\label{sec:structures}
NGC~1291  is a  nearby  galaxy at  8.6  Mpc (mean  redshift-independent
distance from  NED; http://ned.ipac.caltech.edu/). It is  a relatively
massive  (M$_{\star}$ =  5.8$\times$10$^{10} $M$_{\sun}$)  and face-on
galaxy  with an  inclination of  11  degrees when  measured using  the
semi-axes ratio  of the 25.5  mag/arcsec$^2$ isophote at 3.6  $\mu m$.
Recently,  \citet{buta15} morphologically  classified  this galaxy  as
(R)SAB(l,bl,nb)0+.  This  indicates it is a  lenticular, double-barred
galaxy, with an outer ring, a lens, and a barlens.

We performed  a two-dimensional photometric decomposition  of NGC~1291
using the 3.6 $\mu m$ image  provided by the Spitzer Survey of Stellar
Structure  in Galaxies  \citep[S$^4$G,][]{sheth10}.  To  this aim,  we
applied        a        modified       version        of        GASP2D
\citep{mendezabreu08a,mendezabreu17,mendezabreu18} upgraded to account
for  the complex  structural mixing  of this  galaxy.  Details  on the
methodology to  decompose double-barred  galaxies can  be found  in de
Lorenzo-C\'aceres et al.   (in prep.).  We found that  NGC~1291 can be
successfully described  with six  structural components: an  outer and
inner bar modeled with a  Ferrers profile with effective (total) radii
36.7 (131.5)  and 15.5 (29.0)  arcsecs, respectively; a  central bulge
with S\'ersic index $n=2.9$ and effective radius $r_e=9.8$ arcsecs; an
inner exponential  disc surrounding the  inner bar; a lens  related to
the  outer   bar  described   using  a  S\'ersic   surface  brightness
distribution with $n=0.6$; and a faint exponential outer disc.

\begin{figure*}
	\includegraphics[bb= 54 395 558 665,angle=0,width=15cm]{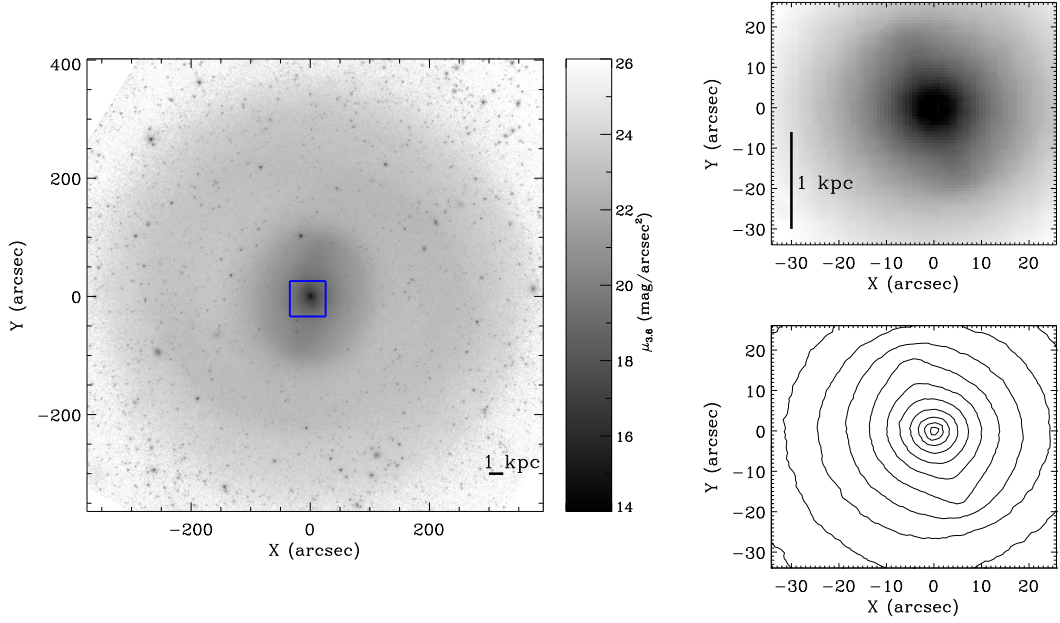}
    \caption{S$^4$G 3.6 $\mu m$ images  of NGC~1291.  Left panel shows
      the      main      outer       structures      described      in
      Sect.~\ref{sec:structures},  i.e., the  outer ring,  outer disc,
      lens,  and  outer bar.   The  blue  square represents  the  MUSE
      $1\arcmin\times1\arcmin$  field of  view.  Right  panels show  a
      zoom-in  (upper   panel)  and   contour  (lower   panel)  images
      highlighting  the central  photometric  structures, i.e.,  inner
      disc,  inner bar,  and central  bulge.  They  correspond to  the
      actual  field of  view  observed with  MUSE,  which is  slightly
      offset with respect to the center  of the galaxy.  North is
        up and East is left.  }
    \label{fig:struc}
\end{figure*}

Bar sizes  for NGC~1291  have  been previously  reported in  the
  literature.  \citet{erwin04}, using measurements based on azimutally
  averaged ellipticity profiles, obtained a  lower (upper) limit of 17
  (24)  and   89  (140)   arcsecs  for  the   inner  and   outer  bar,
  respectively.  \citet{herreraendoqui15}  measured  the  bar  lengths
  using a visual inspection of the S$^4$G images. They found values of
  19.3  and 97.2  for the  inner and  outer bar,  respectively.  Using
  photometric  decompositions, \citet{salo15}  measured a  total outer
  bar size of 158 arcsecs.  Despite  the well known differences in the
  bar size when using different methods \citep{micheldansacwozniak06},
  our values are in relatively good agreement with the literature, and
  they represent the most accurate modelling of the stellar structures
  in NGC~1291 performed  so far.  In addition,  the stellar structure
inventory obtained from the photometric decomposition agrees well with
the visual  classification provided  in \citet{buta15},  and resembles
the    {\it   Russian    Doll}    pattern    described   earlier    by
\citet{devaucouleurs75} as  a repeated lens-bar-nucleus  pattern.  The
main difference is that we do not  find the presence of a barlens, but
an inner  disc surrounding the inner  bar as indicated by  the stellar
kinematic map.  We  do not include, in  our photometric decomposition,
neither the outer ring nor  the {\it ansae} (symmetric enhancements at
the   ends  of   the  stellar   bar)  observed   in  the   inner  bar.
Figure~\ref{fig:struc} shows  a 3.6 $\mu  m$ S$^4$G image  of NGC~1291
where these stellar structures are visible.

\section{TIMER observations of NGC~1291}
\label{sec:observations}

The  TIMER  project is  a  survey  employing the  MUSE  integral-field
spectrograph to study the central $1\arcmin\times1\arcmin$ of a sample
of 24 nearby ($d<40$Mpc) barred  galaxies with nuclear rings and other
central structures such  as nuclear spiral arms, inner  bars, and inner
discs (see Paper I).  One of the project's main goals  is to study the
star formation histories of such  structures to infer the cosmic epoch
of the  formation of the  bar and the  dynamical settling of  the main
disc  of the  host galaxy.   The methodology  was demonstrated  with a
pilot study of NGC~4371 \citep{gadotti15}.

The corresponding  observations of NGC~1291 were  performed during ESO
Period 97 (April to September 2016)  with a  typical seeing on the
  combined  datacube of  $1.1\arcsec$ (spaxel  size =  0.2$\arcsec$),
mean spectral  resolution of  2.65\,\AA (FWHM), and  spectral coverage
from  4750\,\AA\, to  9350\,\AA.  NGC1291,  being one  of the  closest
galaxies in the TIMER  sample, the MUSE $1\arcmin\times1\arcmin$ field
of   view  (FoV)   corresponds  to   a  (almost)   square  region   of
$\approx2.5$\,kpc on a side (i.e., $1\arcsec$ corresponds to 42\,pc).

The  MUSE  pipeline (version  1.6)  was  used  to reduce  the  dataset
\citep{weilbacher12}, correcting  for bias and  applying flat-fielding
and illumination  corrections, as well as  wavelength calibration. The
exposures   were  flux-calibrated   through  the   observation  of   a
spectrophotometric  standard  star,  which  was also  used  to  remove
telluric features. Dedicated empty-sky exposures and a PCA methodology
were employed to  remove signatures from the  sky background. Finally,
the exposures were also finely registered astrometrically, so that the
point  spread function  of the  combined cube  is similar  to that  in
individual  exposures.  More  details  can  be  found  in  Gadotti  et
al. (2018, submitted).

\subsection{Stellar kinematics}

The fully  reduced data  cube was spatially  binned using  the Voronoi
scheme of \citet{cappellaricopin03} to achieve a signal-to-noise ratio
of  approximately 40  per  pixel on  each spatial  bin.   We used  the
penalised     pixel    fitting     (pPXF)     code    developed     by
\citet{cappellariemsellem04}  to extract  the stellar  kinematics from
the binned spectra, masking emission lines and using the E-MILES model
library  of  single  age   and  single  metallicity  populations  from
\citet{vazdekis16}  as stellar  spectral  templates.  Additionally,  a
multiplicative low-order  Legendre polynomial was included  in the fit
to account  for small differences  in the continuum shape  between the
galaxy  and  template spectra.   We  restricted  our analysis  to  the
rest-frame wavelength range between  4750\,\AA\ and 5500\,\AA\ for the
minimisation, after verifying that  including the whole spectral range
produces no noticeable differences in the derived kinematics.

\begin{figure*}
	\includegraphics[bb= 120 0 490 858,angle=90,,width=18.5cm]{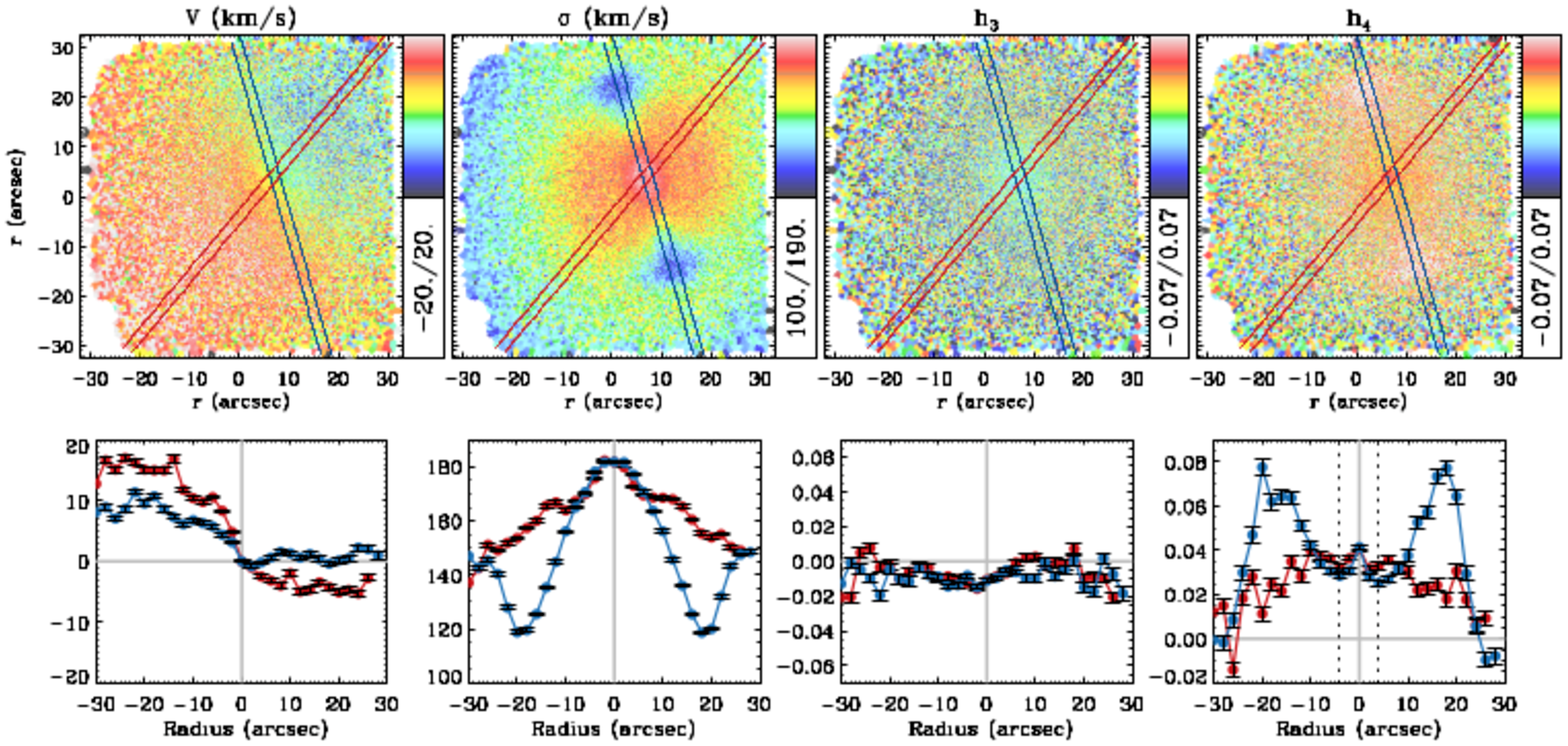}
    \caption{Stellar  kinematic maps  (top rows)  and radial  profiles
      (bottom  rows)  obtained from  the  MUSE  TIMER observations  of
      NGC~1291.  From  left to  right: velocity,  velocity dispersion,
      $h_3$, and  $h_4$ moments.   The radial  profiles in  the bottom
      panels were extracted along 2$\arcsec$ width pseudo-slits (shown
      in the  maps) for both the  inner bar major axis  (blue) and the
      galaxy major axis  (red). The positions of the  $h_4$ minima are
      shown with dashed lines in the bottom right panel. North is
        up and East is left.}
    \label{fig:kin}
\end{figure*}

The  LOSVD was  parameterised with  the mean  stellar velocity  ($V$),
stellar   velocity  dispersion   ($\sigma$),  and   the  Gauss-Hermite
higher-order    moments    $h_3$    and   $h_4$    \citep[see,    {\it
    e.g.},][]{vandermarelfranx93}.  We refer the reader to Paper I for
further details.  The measured stellar kinematic maps for NGC~1291 are
shown in Fig.~\ref{fig:kin}.  We also extracted radial profiles of the
different moments of the LOSVD along  the major axes of both the inner
bar and  inner disc, corresponding  to position  angles of 16  and 140
degrees (North-East),  respectively. The inner disc  position angle is
obtained as  the perpendicular  to the  zero-velocity line  within the
FoV.  We chose  a pseudo-slit width for both profiles  of 2 arcsecs to
account for the MUSE PSF. Errors  shown in the radial profiles account
for both  errors in the individual  spaxels and the standard  error of
the mean within the pseudo-slit.

\section{Results and discussion}
\label{sec:results}

The  rightmost panels  of  Fig.~\ref{fig:kin} show  the $h_4$  spatial
distribution  and  radial  profiles of  NGC~1291.   Clear,  symmetric,
double minima in the $h_4$ are seen  along the major axis of the inner
bar.  These minima are neither observed  along the galaxy nor the main
bar major axis indicating that they are related to the presence of the
inner bar.  The double minima occur at a projected distance of $\sim 4
\arcsec$, which corresponds to 26\% of the inner bar effective radius.
Following these  minima, the radial  profile of $h_4$ along  the major
axis of the bar grows until  it reaches two maxima at radii comparable
to the inner bar radii.  These maxima are not part of a positive $h_4$
ring  around  the  inner  bar,  as it  has  been  found  in  numerical
simulations   \citep{du16},  but   they  nicely   correspond  to   the
$\sigma-$hollows  observed in  the velocity  dispersion profile.   The
$\sigma-$hollows are a  kinematic confirmation for the  presence of an
inner bar as demonstrated  by \citet{delorenzocaceres08}, and are also
not observed along the major axis of the galaxy.

As  demonstrated by  numerical simulations  of single-barred  galaxies
\citep{debattista05,iannuzzi15},  bi-symmetric  minima  in  the  $h_4$
profile along the bar major axis indicate the presence of a vertically
extended B/P structure in the bar.  This kinematic criterion to detect
B/P in  face-on galaxies  has been previously  confirmed in  main bars
\citep{mendezabreu08b,mendezabreu14}, {\em but this  is the first time
  a B/P  is detected in a  face-on inner bar}. Hints of multiple
  B/P   in   edge-on   galaxies    have   also   been   discussed   in
  \citet{ciamburgraham16}.    Remarkably,   these  results   strongly
suggest  that inner  and  outer bars  in  double-barred galaxies  are
analogous systems, governed by the same physical processes.

A closer look  at Fig.~\ref{fig:kin} reveals that the  $h_4$ minima do
not  occur  at  negative  values   of  $h_4$,  as  predicted  in  some
simulations  \citep{debattista05},  but  they   happen  at  $h_4  \sim
0.03$. This  is not surprising  since the inner  bar is embedded  in a
more complex dynamical environment than  the outer (single) bars which
are usually  simulated.  Numerical simulations have  also demonstrated
that the  presence of  a central  dispersion-dominated system  tend to
smooth    the    $h_4$    radial    profile    (see    figure~14    in
\citealt{debattista05}).   We indeed  found photometric  ($n=2.9$) and
kinematic  (velocity dispersion  gradient)  evidence  for a  classical
bulge (i.e., a dispersion-dominated system) in NGC~1291.  In addition,
more  recent  numerical  simulations   including  the  effect  of  gas
dissipation  have shown  that  symmetric double  minima  in the  $h_4$
profile, even if not negative, are  a clear signature for the presence
of   a   B/P  structure   (bottom   right   panel  of   figure~24   in
\citealt{iannuzzi15}).

An  alternative configuration  able to  mimic the  observed $h_4$  and
$\sigma$ maps  would invoke  the presence  of a  counter-rotating disc
instead of the inner bar.  Nevertheless,  if the disc lied in the same
equilibrium  plane  of  the  galaxy,   we  would  expect  a  ring-like
maximum/minimum in the $h_4$/$\sigma$ maps,  which is not observed. On
the  other  hand,  if  the  disc  was  settled  perpendicular  to  the
line-of-sight (LOS) and aligned to the  inner bar major axis, then its
higher LOS  velocities would stand out  in the velocity map,  which is
not  observed  either.   Therefore  we  consider  unlikely  these  two
possibilities, reinforcing the  conclusion of a B/P  associated to the
inner bar.

The presence of  a B/P structure in the inner  bar of NGC~1291 imposes
key constraints on the origin and evolution of double-barred galaxies.
First,  the  evidence  that  inner  bars  can  also  develop  vertical
structures such as  B/P, and therefore they also  suffer from vertical
instabilities, suggests  they are  more similar  to outer  (main) bars
than previously thought.   Our results indicate that  inner bars might
mimic the evolution of outer bars  with one (or perhaps more) buckling
phases that must be reproduced  by numerical simulations.  To our best
knowledge  there is  no simulation  of double-barred  galaxies showing
that inner bars undergo a  buckling phase, but former orbital analysis
have  found  vertically  extended   periodic  orbit  families  in  the
innermost  parts  of  bars \citep{skokos02,patsis02}.   Secondly,  the
presence of  a B/P  introduces a  timing constraint  on the  inner bar
formation.

Although the formation timescales of both  bars and B/Ps depend on the
specific  characteristics of  the  numerical simulations  such as  gas
fraction,  halo  shape,  and  bulge dominance,  the  general  timeline
includes  the  formation of  the  bar  itself  ($\sim1$ Gyr)  and  the
consequent formation of  the B/P, after the bar is  formed, due to the
first buckling  phase ($\sim1$ Gyr).    Under the  assumption that
  the evolutionary  timescales of  both the inner  and outer  bars are
  similar, we speculate that the inner  bar of NGC~1291 should have an
  age of  at least  $\sim2$ Gyrs,  which is  longer than  the lifetime
  expected   for   inner   bars    in   some   numerical   simulations
  \citep{friedlimartinet93}.    However,    further   simulations   of
  double-barred galaxies (showing the  B/P formation) are necessary to
  better  understand  their  time  evolution.  In  particular,  it  is
  possible that inner bars might form in much shorter times than outer
  bars \citep[$\sim$150  Myrs;][ but  other authors show  more similar
    timescales                     $\sim$0.5-1                    Gyrs;
    \citealt{du15,wozniak15}]{debattistashen07}.   Notwithstanding the
  previous  caveats, we  found independent  support to  our conclusion
  that  inner bars  are  long-lived  through the  typical  age of  the
  stellar populations present in the  central parts of NGC~1291, which
  turns out to be $\sim$ 6-7  Gyrs (discussed in de
  Lorenzo-Caceres et al.  2018, submitted).

\section{Conclusions}
\label{sec:conclusions}

We used the superb quality of the MUSE TIMER data to identify, for the
first  time, a  B/P  structure in  the inner  bar  of a  double-barred
galaxy, NGC~1291.  At the  location of the  B/P, the  vertical density
distribution of  stars becomes broader and  numerical simulations have
demonstrated  that  this  effect  can  be  observationally  traced  as
symmetric  double minima  in the  radial profile  of the  fourth-order
Gauss-Hermite moment  ($h_4$) of the  LOSVD along the bar  major axis.
This kinematic  diagnostic was confirmed  in the single bar  of NGC~98
\citep{mendezabreu08b} and in  this paper we show that  inner bars can
also buckle, as we detect the  same kinematic feature in the inner bar
of NGC~1291.

The  presence of  B/P structures  in the  inner bars  of double-barred
galaxies suggests that they follow the same evolutionary path as outer
bars,  which   have  been  extensively  studied   in  the  literature.
Therefore,  inner  bars  should  grow  both  radially  and  vertically
depending on  the interaction with the outer bar and the amount of
  angular momentum  exchanged by  the latter  with other  baryonic and
  dark matter components, following buckling phases as those predicted
  for  single  barred  galaxies.   In addition,  we  claim  that  the
presence of  a B/P  in the inner  bar, combined with  the ages  of the
stellar populations (presented in de  Lorenzo-Caceres et al.  2018, in
prep.), suggests that inner bars  can be long-lived structures lasting
several  Gyrs.  Our  results rule  out inner  bar formation  scenarios
where they  are short-lived structures,  like those invoking  an inner
gaseous bar as the progenitor of the inner stellar bar.

Additional  detections  of B/P  structures  in  inner bars,  as  those
reported in this letter, for a  sample of double barred galaxies would
allow  to  further constrain  their  formation  models.  This  is  now
feasible with integral-field spectrographs such as MUSE.

\section*{Acknowledgements}

We thank the  anonymous referee and the scientific  editor for her/his
many valuable comments  which helped to improve this  paper.  AdLC and
JMA thank  ESO for the  warm hospitality  during a science  visit when
part of this work was done.  JMA acknowledges support from the Spanish
Ministerio  de  Economia  y   Competitividad  (MINECO)  by  the  grant
AYA2017-83204-P.  AdLC  and J.   F-B acknowledges support  from MINECO
grant  AYA2016-77237-C3-1-P.   GvdV   acknowledges  funding  from  the
European  Research  Council  (ERC)  under grant  agreement  No  724857
(Consolidator Grant ArcheoDyn).  PSB  acknowledges support from MINECO
grant  AYA2016-77237-C3-2-P. Based  on observations  collected at  the
European  Organisation  for  Astronomical  Research  in  the  Southern
Hemisphere under ESO programme 097.B-0640(A).




\bibliographystyle{mnras}
\bibliography{reference} 








\bsp	
\label{lastpage}
\end{document}